\documentclass[prd,twocolumn,nofootinbib]{revtex4-2}
\usepackage{amsfonts}
\usepackage{graphicx}
\usepackage{xspace}
\usepackage{comment}
\usepackage{dcolumn}
\usepackage{amsmath,bm,amssymb}
\usepackage{psfrag}
\usepackage[caption=false]{subfig}
\usepackage{braket}
\usepackage{slashed}
\usepackage{color,xcolor,soul}
\definecolor{urlblue}{rgb}{0.2,0.4,0.7}
\definecolor{citegreen}{rgb}{0,0.6,0.2}
\definecolor{linkred}{rgb}{0.9,0.2,0.1}
\usepackage{hyperref}
\hypersetup{
colorlinks=true, citecolor=citegreen, linkcolor=blue,urlcolor=urlblue}
\def\l{\left}
\def\r{\right}

\def\eqv{\equiv}
\def\nn{\nonumber}

\def\fr{\frac}

\def\gst{\tilde g_{\rm S}}
\def\as{\alpha_{\rm S}}
\def\cg{c_\Gamma}
\def\cgt{\tilde c_\Gamma}

\def\eps{\epsilon}
\def\veps{\varepsilon}

\def\ra{\rightarrow}

\def\Amp{\mathcal{M}}
\def\as{\alpha_{\rm S}}
\def\beq{\begin{equation}}
\def\eeq{\end{equation}}
\def\beeq{\begin{eqnarray}}
\def\eeeq{\end{eqnarray}}
\def\nn{\nonumber}
\def\eq#1{Eq.~(\ref{#1})}

\def\ra{\rangle}

\def\ket#1{|{#1}\ra}

\def\as{\alpha_{\mathrm{S}}}

\def\wp{\widetilde p}

\def\bom#1{{\mbox{\boldmath $#1$}}}
\def\sp{\bom{Sp}}
\def\js{\bom{J}}

\def\eps{\epsilon}

\allowdisplaybreaks

\begin{document}

\title{Catani's generalization of collinear factorization breaking}

\author{Leandro Cieri}
\email{lcieri@ific.uv.es}
\author{Prasanna K. Dhani}
\email{dhani@ific.uv.es, prasanna.dhani@physik.uzh.ch}
\author{Germ\'{a}n Rodrigo}%
\email{german.rodrigo@csic.es}
\affiliation{%
Instituto de F\'{\i}sica Corpuscular, 
Universitat de Val\`encia - Consejo Superior de Investigaciones Cient\'{\i}ficas, E-46980 Paterna, Valencia, Spain
}%
\begin{abstract}
We consider the most general form of soft and collinear factorization for hard-scattering amplitudes to all orders in perturbative quantum chromodynamics. Specifically, we present the generalization of collinear factorization to configurations with several collinear directions, where the most singular behavior is encoded by generalized collinear splitting amplitudes that manifestly embed the breaking of strict collinear factorization in spacelike collinear configurations. We also extend the analysis to the simultaneous soft-collinear factorization with multiple collinear directions where naive multiplicative factorization does not hold. As an illustrative example of factorization breaking, we present explicit results at the one-loop level in the soft-collinear limit.
\end{abstract}

\maketitle

\section{Introduction}
\label{sec:intro}
\begin{figure}[t]
\centering
\includegraphics[width=\linewidth]{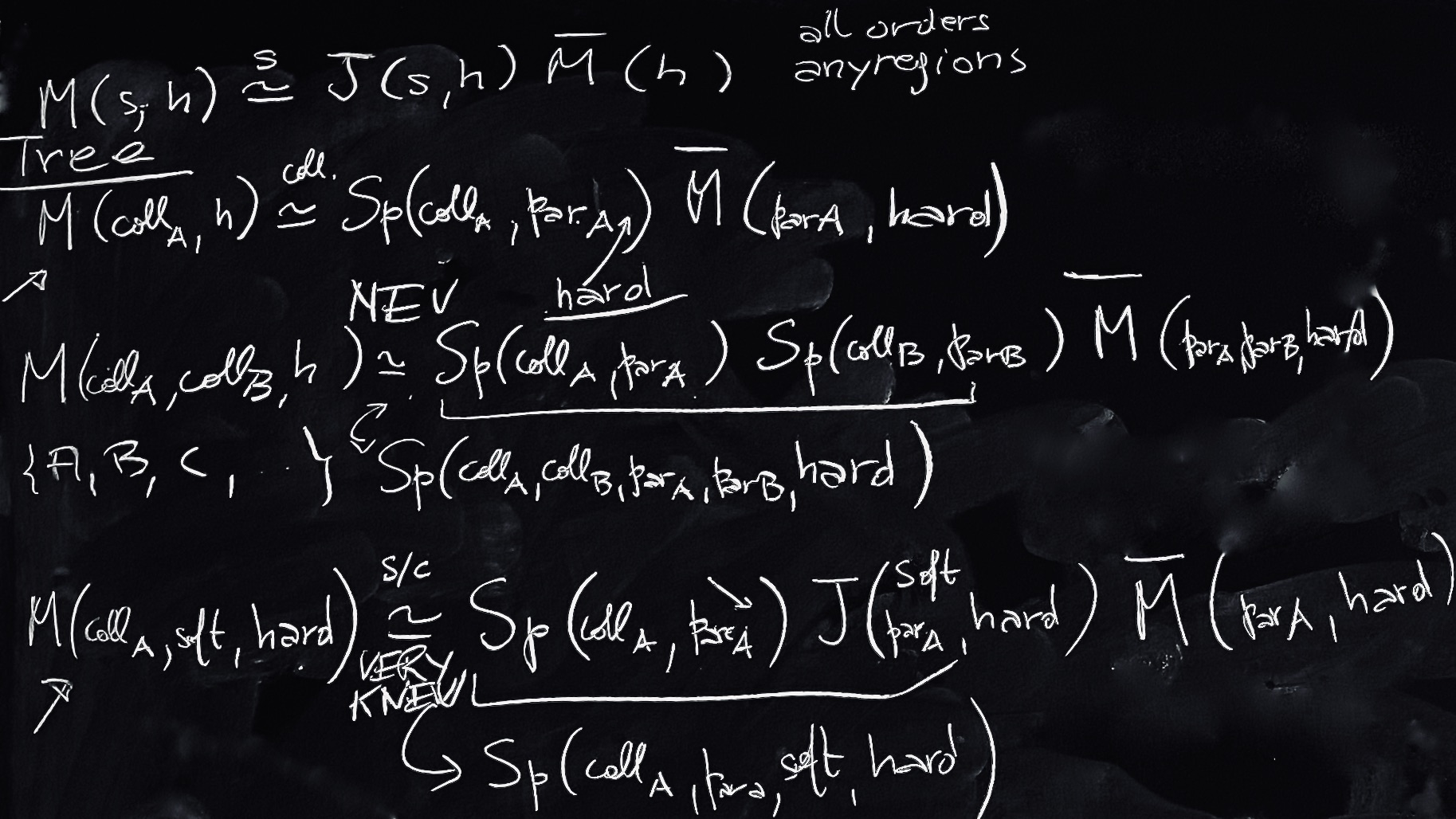}
\caption{Footage of original handwritten blackboard discussion by Stefano Catani, September 2023, Sesto Fiorentino, Firenze, Italy. \label{fig:handwritting}}
\end{figure}

Hard-scattering amplitudes in perturbative quantum chromodynamics (QCD) are singular if the energy of one or more partons, quarks and gluons is negligibly small and/or if two or more parton momenta are parallel to each other. The singular behavior in these kinematic configurations is factorized~\cite{Collins:1989gx,Ellis:1996mzs} and encoded in process-independent factors such as collinear splitting amplitudes and soft currents. Based on the lowest-order versions of soft and collinear factorization~\cite{Altarelli:1977zs,Bassetto:1983mvz}, general algorithms for predicting jet cross sections at next-to-leading order~(NLO) were proposed in the famous works in Refs.~\cite{Frixione:1995ms,Catani:1996vz,Frixione:1997np,Catani:2002hc}. In the last twenty years, tremendous efforts have been made to advance to next-to-NLO (NNLO) and next-to-NNLO (N$^3$LO), and there are now a variety of different methodological approaches (see, e.g., the reviews in Refs.~\cite{Heinrich:2020ybq,TorresBobadilla:2020ekr,Agarwal:2021ais}) capable of achieving high perturbative orders. Also, resummed calculations to all perturbative orders have achieved next-to-next-to-next-to-leading logarithmic accuracy in some specific cases (see, Refs.~\cite{Camarda:2021ict,Billis:2021ecs,Neumann:2022lft} and references therein). All these advances rely on the detailed knowledge of the underlying QCD factorization properties of hard-scattering amplitudes in the soft and collinear limits.
At $\mathcal{O}(\as^2)$, soft factorization at tree level is studied in Refs.~\cite{Berends:1988zn,Campbell:1997hg,Catani:1999ss} and at one loop with single soft emission in Refs.~\cite{Bern:1995ix,Bern:1998sc,Bern:1999ry,Catani:2000pi,Bierenbaum:2011gg}. At the next perturbative order, i.e., $\mathcal{O}(\as^3)$, the triple-soft limit of hard-scattering amplitudes at tree level is calculated in Refs.~\cite{Catani:2019nqv,DelDuca:2022noh,Catani:2022hkb}, the double-soft limit of one-loop amplitudes is computed in Refs.~\cite{Zhu:2020ftr,Catani:2021kcy,Czakon:2022dwk}, and the two-loop case with single soft emission is analyzed in Refs.~\cite{Li:2013lsa,Duhr:2013msa,Dixon:2019lnw}. Furthermore, at $\mathcal{O}(\as^4)$ a few explicit results are recently presented in Refs.~\cite{Chen:2023hmk,Herzog:2023sgb,Chen:2024hvp}.
Collinear factorization is considered universal and process independent for theory predictions at high-energy colliders. Strictly speaking, the singular factors arising in scattering amplitudes from collinear configurations are expected to depend only on the momenta and quantum numbers of the external partons involved in the collinear splitting process, and this aspect is known in the literature as strict collinear factorization (SCF). At $\mathcal{O}(\as^2)$, tree-level collinear splitting processes are studied in Refs.~\cite{Campbell:1997hg,Catani:1998nv,Catani:1999ss,Dhani:2023uxu,Craft:2023aew}, while double-collinear splitting processes at one loop are analyzed in Refs.~\cite{Bern:1998sc,Kosower:1999rx,Bern:1999ry,Catani:2011st,Sborlini:2013jba}. At the next perturbative order, i.e., $\mathcal{O}(\as^3)$, the quadruple-collinear limit of hard-scattering amplitudes is calculated in Refs.~\cite{DelDuca:1999iql,Birthwright:2005ak,Birthwright:2005vi,DelDuca:2019ggv,DelDuca:2020vst}, while the triple-collinear limit of one-loop amplitudes is discussed in Refs.~\cite{Catani:2003vu,Catani:2011st,Sborlini:2014mpa,Sborlini:2014kla,Badger:2015cxa,Czakon:2022fqi}, and the two-loop case with double-collinear emission is analyzed in Refs.~\cite{Bern:2004cz,Badger:2004uk,Duhr:2014nda}. At $\mathcal{O}(\as^4)$, double-collinear splitting processes at three loops are recently presented in Ref.~\cite{Guan:2024hlf}.
A seminal paper~\cite{Catani:2011st} led by Stefano Catani showed, however, that SCF is not valid beyond the tree level in perturbation theory. In particular, SCF remains true for configurations where all collinear partons are produced in the physical final state termed as timelike (TL) collinear but it is broken for other kinematic collinear configurations generally denoted as spacelike (SL) collinear. This effect has a purely quantum field theory origin from Feynman loop diagrams and it is related to the intriguing color coherence properties of QCD. In particular, the soft modes in the loop couple the collinear partons in the final state to noncollinear ones in the initial state that are far apart in phase-space kinematics. As a consequence, in the SL collinear region, QCD scattering amplitudes fulfil a generalized factorization, in which the singular factors retain certain dependence on the momenta and quantum numbers of the noncollinear external partons.
The SCF breaking effects have more general implications in the context of perturbative QCD predictions for jet production at colliders. Starting from N$^3$LO, these effects may interfere with the cancellation mechanism of infrared~(IR) divergences that leads to the factorization theorem of mass (collinear) singularities. Besides challenging the process independent validity of the age-old mass-singularity factorization theorem, SCF breaking effects also emerge in the context of factorization of transverse-momentum dependent distributions~\cite{Rogers:2010dm,Echevarria:2011epo,Echevarria:2012js,Rogers:2013zha}. The resummation of large logarithmic terms produced by collinear parton evolution is affected by the breaking of SCF: parton evolution gets tangled up with the color and kinematic structure of the hard-scattering subprocess, and this leads to the appearance of entangled large logarithms. Indeed, the physical mechanism that generates the breaking of SCF is directly related to the mechanism that produces superleading logarithms~\cite{Forshaw:2006fk,Forshaw:2008cq,Keates:2009dn,Becher:2021zkk,Becher:2024nqc} in jet cross sections. A recent study of the intricacies of the soft photon theorem in QCD can be found in Ref.~\cite{Ma:2023gir}.
{\it The purpose of this Letter is to disseminate Stefano Catani's ideas on the generalization of soft and collinear factorization that apply to both TL and SL domains to all orders in QCD perturbation theory. In particular, we cherish the memory of our last blackboard conversation (Fig.~\ref{fig:handwritting}) we had with him in September 2023 at the Department of Physics in Sesto Fiorentino, Firenze, Italy.}

\section{QCD Factorization in IR limits}
\label{sec:IR_factorization}
Let us consider a generic scattering process with $n$ external partons, massless quarks and gluons, together with an arbitrary number of non-QCD particles, particles that are color singlet, such as Higgs bosons, vector bosons, leptons, photons, etc. The scattering amplitude for this process is denoted by $\Amp(p_1,\ldots,p_n)$, where $p_i$ is the momentum of the parton with flavor $a_i\, (a_i = g, q\, {\rm or}\, \bar{q})$, while the dependence on non-QCD particles is implicitly understood. The external QCD particles are on shell $(p_i^2=0)$ and have physical spin polarizations. Here, the scattering amplitude is defined in an arbitrary number of space-time dimensions.
In the context of QCD perturbation theory, the scattering amplitude $\Amp(p_1,\ldots,p_n)$ is expanded to higher loop orders as follows:
\begin{align}
    \label{eq:M_pert_exp}
    \Amp(p_1,\ldots,p_n) = \sum_{i=0} \Amp^{(i)}(p_1,\ldots,p_n)~,
\end{align}
where $\Amp^{(0)}$ denotes the lowest-order contribution, $\Amp^{(1)}$ denotes the one-loop contribution, and so on. 
We will consider kinematic configurations where subsets of the external particles become soft or collinear. To simplify the presentation, we will denote by $s$ the subset of soft partons and their momenta, by $c_I$ a subset of collinear partons, where the subindex $I$ could label collinear subsets in different collinear directions, and by $h$ the remaining hard partons.
\subsection{Soft factorization}
\label{sec:soft_factorization}
We are interested in the behavior of a hard-scattering amplitude $\Amp$ in the kinematic configuration where $m$ massless external partons become soft. This limit is formally obtained by first scaling all the momenta of the soft partons by a factor $\lambda$
and then considering the limit $\lambda \rightarrow 0$. In this limit, the leading singular behavior is captured as follows:
\begin{align}
    \label{eq:soft_behaviour}
    \left. \Amp(\lambda s, h)\right|_{\lambda\to 0 } \sim \frac{1}{\lambda^m}{\rm mod}\,(\ln^r \lambda) + \cdots~.
\end{align}
The powerlike behavior $\lambda^{-m}$ on the right-hand side in \eq{eq:soft_behaviour} determines the dominant singular terms of $\Amp$, the logarithmic factor $\ln^r \lambda$ indicates a break in the naive scaling behavior due to higher-order quantum loop  contributions, and the dots denote the subdominant singular behavior of relative suppression $\mathcal{O}(\sqrt{\lambda})$ at least. 
The emission of soft partons has no influence on the kinematics, spins and flavors of the hard partons present in the scattering process. However, they influence the color structure, since the emission of a soft parton always carries color charges and these do not depend on its softness. As a result, in QCD, there are color correlations, and the scattering amplitude is not fully factorized, in contrast to Abelian theories such as quantum electrodynamics, where the contribution of soft photons factorizes from the rest of the scattering amplitude.
In this multiparticle soft limit, the dominant singular behavior of $\Amp(s,h)$ is expressed by the following process-independent factorization ~\cite{Campbell:1997hg,Bern:1995ix,Bern:1999ry,Catani:1999ss,Catani:2000pi,Feige:2014wja}
\begin{align}
\label{eq:soft_factorization}
    |\Amp(s,h)\rangle  \simeq \js(s;h) |\Amp(h)\rangle\,,
\end{align}
where the factor $\js(s;h)$ on the right-hand side of \eq{eq:soft_factorization}
is the multiparticle soft current that embodies the dominant singular behavior, the reduced scattering amplitude $|\Amp(h)\rangle$ is simply obtained from the full amplitude on the left-hand side by removing the $m$ external legs with soft parton momenta, and the symbol $\simeq$ indicates that the subdominant singular terms are neglected.
It is important to note that the soft current $\js(s;h)$ depends on both soft and hard partons through their momenta and quantum numbers, such as color. We also note that the soft current is proportional to the unit operator in the spin subspace of the hard partons. Despite its dependence on the hard partons, the soft current is universal, i.e., it does not depend on the specific hard-scattering amplitude and the corresponding specific physical process.
The soft current $\boldsymbol{J}$ in \eq{eq:soft_factorization} can be expressed in perturbative QCD, by a loop expansion analogous to~\eq{eq:M_pert_exp},
\begin{align}
    \label{eq:SC_loop_exp}
    \js(s;h) = \sum_{i=0} \js^{(i)}(s;h)~,
\end{align}
where $\js^{(0)}$ is the tree-level soft current, $\js^{(1)}$ is the one-loop soft current, and so forth.
\subsection{Collinear factorization}
\label{sec:collinear_factorization}
Scattering amplitudes in QCD also factorize when two or more external on-shell momenta become collinear to each other. Collinear factorization and the corresponding splitting processes are classified into two categories: TL and SL, as discussed above.
We are interested in the behavior of the scattering amplitude $\Amp$ in the kinematic configuration in which $m$ of the $n$ external partons become parallel to each other. Without loss of generality, we assume that the collinear momenta are $p_1,\ldots,p_m$ and define their sum as $p_{1\ldots m}$. Since the decaying parent particle is generally off shell $(s_{1\ldots m}\equiv p_{1\ldots m}^2\neq 0)$, we define an on shell vector
\begin{align}
    \label{eq:onshellvec}
    \wp^{\mu} = p_{1\ldots m}^{\mu}-\frac{s_{1\ldots m}}{2n\cdot p_{1\ldots m}}n^{\mu},
\end{align}
such that it fulfils $\wp^2=0$. Then we introduce the Sudakov parametrization for the collinear momenta~\cite{Catani:1999ss,Catani:2011st} as follows:
\begin{align}
    p_i^{\mu} = x_i \widetilde{p}^{\mu} + k_{\perp i}^{\mu} - \frac{k_{\perp i}^2}{x_i}\frac{n^{\mu}}{2\,n\cdot \widetilde{p}}\,, ~ i \in c\equiv\{1,\ldots,m\},
\end{align}
where the vector $\wp^\mu$ denotes the collinear direction and the lightlike auxiliary vector $n^{\mu}$ ($n^2=0$) is necessary to specify how the collinear direction is approached or, equivalently, to specify the transverse components $k_{\perp i}$ ($k_{\perp i}\cdot \wp = k_{\perp i}\cdot n= 0$, with $k_{\perp i}^2<0$).
Now, the $m$-parton collinear limit is precisely defined by scaling the transverse momenta $k_{\perp i}^{\mu}$ with an overall factor $\lambda$, and then setting the limit $\lambda\rightarrow 0$. If we keep the most singular terms in $\lambda$ and neglect subdominant terms, we obtain the following amplitude-level factorization~\cite{Catani:2011st}
\begin{align}
\label{eq:TLcoll_factorization}
    \ket{\Amp(c, h)} \simeq 
    \sp(c; \wp) \, 
    \ket{\Amp(\wp, h)}~.
\end{align}
The factorization in \eq{eq:TLcoll_factorization} relates the original scattering amplitude with $n$ partons on the left-hand side to a reduced scattering amplitude with $n-m+1$ partons on the right-hand side. The latter is obtained by replacing the $m$ collinear partons $a_1,\ldots,a_m$ by a single parent parton whose momentum is $\wp^\mu$ and whose flavor $a$ is determined by flavor conservation.
The process dependence of \eq{eq:TLcoll_factorization} is fully taken into account in the amplitudes on both sides. The factor $\sp(c; \wp)$ is called the splitting amplitude, which encodes the dominant singular behavior in the collinear limit and depends only on the momenta and quantum numbers such as color, flavor and spin of the partons involved in the collinear splitting process $a\rightarrow a_1+\cdots+a_m$. This property of collinear factorization is also referred to in the literature as SCF.
As found in a highly regarded paper~\cite{Catani:2011st} by Catani {\it et al.}~(see also Refs.~\cite{Forshaw:2012bi,Sterman:2022gyf}), SCF is only valid in the so-called TL collinear region and breaks down for loop amplitudes in the SL collinear region. In the SL collinear region, the splitting amplitude depends on both the collinear and noncollinear partons through their momenta and quantum numbers. The modified factorization valid for both TL and SL regions is as follows~\cite{Catani:2011st}:
\begin{align}
\label{eq:SLcoll_factorization}
    \ket{\Amp(c, h)} \simeq 
    \sp(c; \wp; h) \, 
    \ket{\Amp(\wp, h)}~. 
\end{align}
Analogous to the soft current in \eq{eq:SC_loop_exp}, the splitting amplitude $\sp$ can be predicted in perturbative QCD and its loop expansion is as follows:
\begin{align}
\label{eq:SM_loop_exp}
    \sp(c;\wp;h) = \sp^{(0)}(c;\wp) + \sum_{i=1}\sp^{(i)}(c;\wp;h)~,
\end{align}
where $\sp^{(0)}$ is the splitting amplitude at tree level, $\sp^{(1)}$ is the one-loop splitting amplitude, and so on. The splitting amplitude at tree level, $\sp^{(0)}$, fulfils SCF and is independent of the hard partons both in the TL and SL regions.
\subsection{Soft-collinear factorization}
\label{sec:soft-collinear_factorization}
We now analyze the scattering amplitude $\Amp(s,c,h)$ for the limiting case in which a subset of partons become soft and at the same time another subset of partons become collinear to each other. We also consider the special case where a subset of collinear partons are soft. We focus for now on the TL collinear limit, i.e., all collinear particles are produced in the physical final state. When examining this soft-collinear limit, we  neglect soft configurations that are uniformly of $\mathcal{O}(p_i)$ when $p_i\rightarrow 0$ and collinear configurations that are uniformly of $\mathcal{O}(k_{\perp i})$ when $k_{\perp i}\rightarrow 0$. Their contributions are accounted for on the basis of the soft and collinear factorizations described in Secs.~\ref{sec:soft_factorization} and \ref{sec:collinear_factorization}. Analogous to Eqs. (\ref{eq:soft_factorization}) and (\ref{eq:TLcoll_factorization}), the singular behavior of the hard-scattering amplitude is factorized in this multipartonic soft-collinear limit and expressed by the following factorization~\cite{Catani:1999ss}:
\begin{align}
    \label{eq:TLsoft-collinear_factorization}
    |\Amp(s,c,h)\rangle \simeq \sp(c;\wp) \, \js(s;\wp;h)  \,|\Amp(\wp, h)\rangle~,
\end{align}
and when some of the collinear partons are soft, the Eq.~(\ref{eq:TLsoft-collinear_factorization}) reduces to
\begin{align}
    \label{eq:TLsoft-collinear_factorization_spl}
    |\Amp(s\subset c,h)\rangle \simeq \sp(s\subset c;\wp)  \,|\Amp(\wp, h)\rangle~,
\end{align}
where the perturbative loop expansion for the splitting amplitude $\sp$ is identical to \eq{eq:SM_loop_exp} without the dependence on the hard noncollinear partons within the argument. The loop expansion for the soft current $\js(s;\wp;h)$ is analogous to \eq{eq:SC_loop_exp}. Note that the effective interaction of the soft current with the collinear partons occurs only through the parent parton due to color coherence. In the following section, we extend the factorization described in Secs.~\ref{sec:collinear_factorization} and \ref{sec:soft-collinear_factorization} to the most general kinematics.
\section{Generalized collinear and soft-collinear factorization}
\label{sec:gen_col_soft-coll}
In this section, we disseminate  Stefano Catani's new ideas on the generalization of QCD factorization of hard-scattering amplitudes in the collinear and simultaneous soft-collinear limits. The generalization of collinear factorization in \eq{eq:TLcoll_factorization}, to two sets of collinear partons, denoted by $c_A$ and $c_B$, with parent momenta $\wp_A$ and $\wp_B$, respectively, is as follows:
\begin{figure}[t]
\centering
\includegraphics[width=.9 \linewidth]{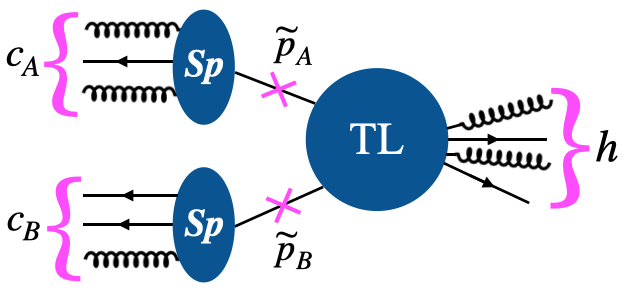}
\caption{TL collinear factorization with two collinear directions.
\label{fig:TL}}
\end{figure}
\begin{figure}[t]
\centering
\includegraphics[width=.9 \linewidth]{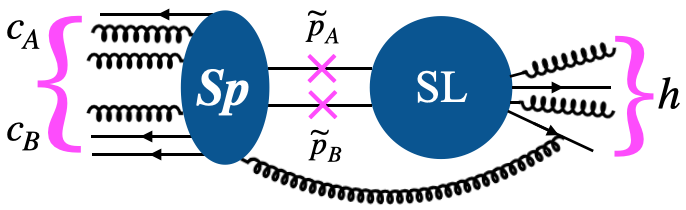}
\caption{SL collinear factorization with two collinear directions.
\label{fig:SL}}
\end{figure}
\begin{align}
   \label{eq:TLcoll_fact_2sets}
   \ket{\Amp(c_A,c_B,h)} \simeq \sp(c_A; \wp_A)\, \sp(c_B; \wp_B)\,
   \ket{\Amp(\wp_A,\wp_B,h)}\,.
\end{align}
Strictly speaking, \eq{eq:TLcoll_fact_2sets} is valid to all orders in perturbation theory only in the TL collinear region (see Fig.~\ref{fig:TL}). However, as discussed in Sec.~\ref{sec:collinear_factorization}, collinear factorization must be modified for a general kinematics that includes the SL collinear region~\cite{Catani:2011st}. The generalization of collinear factorization to more than one collinear direction, e.g., two collinear directions, does not correspond to the naive multiplicative expectation, i.e., the product of two splitting amplitudes encoding each collinear splitting according to \eq{eq:SLcoll_factorization}. Rather, it is given by a single nonfactorizable splitting amplitude acting on the reduced hard-scattering amplitude. The generalized factorization is as follows:
\begin{align}
   \label{eq:SLcoll_fact_2sets}
   \ket{\Amp(c_A,c_B, h)} \simeq \sp(c_A,c_B; \wp_A,\wp_B; h) \, \ket{\Amp(\wp_A,\wp_B,h)}
\end{align}
and is illustrated in Fig.~\ref{fig:SL}. Recently, the first evidence of this generalized factorization was observed in Ref.~\cite{Duhr:2025lyg} in ${\cal N}=4$ super Yang-Mills theory. The collinear factorization in~\eq{eq:SLcoll_fact_2sets} can now easily be extended to any number of collinear directions. Analogous to \eq{eq:SM_loop_exp}, the splitting amplitude $\sp(c_A,c_B; \wp_A,\wp_B; h)$ has a perturbative expansion, where the splitting amplitude at the tree level fulfills SCF
\begin{align}
\label{eq:SM_loop_exp_two-coll-dir}
    \sp^{(0)}(c_A,c_B; \wp_A,\wp_B) &= \sp^{(0)}(c_A; \wp_A)\sp^{(0)}(c_B;\wp_B)~.
\end{align}
Based on~\eq{eq:TLsoft-collinear_factorization}, soft-collinear factorization for two sets of collinear partons and a set of soft partons in the TL collinear region reads as
\begin{align}
    \label{eq:TLsoft-collinear_factorization_compact}
    |\Amp(s,c_A,c_B,h)\rangle & \simeq \sp(c_A;\wp_A) \, \sp(c_B;\wp_B) \, \nn \\ & \times \js(s;\wp_A,\wp_B;h)|\Amp(\wp_A, \wp_B, h)\rangle~,
\end{align}
where the soft current effectively interacts with the parent partons of the two collinear splittings (see, Fig.~\ref{fig:TL soft-collinear}). However, for a general kinematics that includes the SL collinear region, the strict factorization in~\eq{eq:TLsoft-collinear_factorization} breaks down and the new soft-collinear factorization is as follows:
\begin{figure}[t]
\centering
\includegraphics[width=\linewidth]{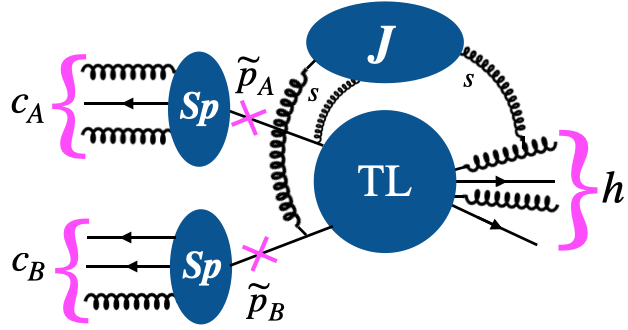}
\caption{TL soft-collinear factorization. The soft current effectively interacts with the collinear parent partons.
\label{fig:TL soft-collinear}}
\end{figure}
\begin{align}
    \label{eq:SLsoft-collinear_factorization}
    |\Amp(s,c,h)\rangle \simeq \sp(c;\wp;s;h)  \,|\Amp(\wp, h)\rangle\,,
\end{align}
where $\sp(c;\wp;s;h)$ is the generalized soft-collinear splitting amplitude, where again only the tree-level component of its perturbative expansion is strictly factorized
\begin{align}
\label{eq:SLsoft-collinear_SA_exp}
    \sp^{(0)}(c;\wp;s;h) &= \sp^{(0)}(c; \wp)\,\js^{(0)}(s;\wp;h)~.
\end{align}
If some of the collinear partons are soft, the Eq.~(\ref{eq:SLsoft-collinear_factorization}) reduces to
\begin{align}
    \label{eq:SLsoft-collinear_factorization-Spl}
    |\Amp(s\subset c,h)\rangle \simeq \sp(s\subset c;\wp;h)  \,|\Amp(\wp, h)\rangle\,.
\end{align}
\begin{figure}
\centering
\includegraphics[width=\linewidth]{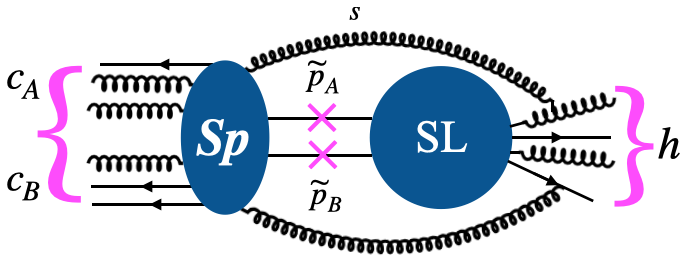}
\caption{SL soft-collinear factorization. The generalized splitting amplitude embeds the soft interactions.
\label{fig:SL soft-collinear}}
\end{figure}

The soft-collinear factorization in~\eq{eq:SLsoft-collinear_factorization} can now easily be generalized to any number of collinear directions. For example, in the case of two collinear directions (see Fig.~\ref{fig:SL soft-collinear}), we have
\begin{align}
   \label{eq:SLsoftcoll_fact_2sets}
   \ket{\Amp(s,c_A,c_B, h)} & \simeq \sp(c_A,c_B; \wp_A,\wp_B; s; h) \, \nn \\ & \times \ket{\Amp(\wp_A,\wp_B,h)}~,
\end{align}
where
\begin{align}
\label{eq:SLsoftcoll_fact_2sets_tree}
   \sp^{(0)}(c_A,c_B; \wp_A,\wp_B; s; h) &= \sp^{(0)}(c_A; \wp_A)\sp^{(0)}(c_B;\wp_B) \nn \\ & \times \js^{(0)}(s;\wp_A,\wp_B;h)~.
\end{align}
The generalized splitting amplitude in \eq{eq:SLsoftcoll_fact_2sets} embeds the soft interactions and retains some dependence on the hard partons.

For an illustrative example of Eq.~(\ref{eq:SLsoft-collinear_factorization-Spl}), we consider an independent calculation of the double soft-gluon current at one-loop in QCD~\cite{Zhu:2020ftr,Czakon:2022dwk,Cieri:2025xxx} and investigate the triple-collinear limit $(p_1\parallel q_1\parallel q_2)$, with $q_1$ and $q_2$ soft, in both TL ($p_1$ in the physical final state) and SL ($p_1$ and an additional hard parton $p_2$ both in the physical initial state) regions. We find the factorization breaking (FB) contribution
\begin{align}
\label{eq:j12gFBR}
&\Delta\sp^{(1)a_1a_2}_{\rm FB}(q_1,q_2\subset c;\wp;h) = \gst^4 f^{a_1 bd}f^{a_2 cd}\varepsilon_1^{\alpha_1}(q_1)\varepsilon_2^{\alpha_2}(q_2)
\nn\\ 
& \times \left\{\bom{T}_1^b\bom{T}_2^c F_{\alpha_1\alpha_2}(q_1,q_2)+ \bom{T}_2^b\bom{T}_1^c F_{\alpha_2\alpha_1}(q_2,q_1) \right\},
\end{align}
where
\begin{align}
\label{eq:Falp1alp2}
& F_{\alpha_1\alpha_2}(q_1,q_2) = \cg e^{\imath\pi\eps}(2q_1\cdot q_2)^{-\eps}
\nn\\ 
& \times \bigg[\left(\frac{2\imath\pi}{\eps}+2\pi^2\right) J^{(0)}_{\alpha_1\alpha_2}(q_1,q_2) + \imath\pi \left( \cdots\right) + {\cal O}(\eps)\bigg].
\end{align}
For notational details, we refer interested readers to App.~\ref{sec:details}. To the best of our knowledge, this is the first instance where a FB term $\propto \pi^2$ at ${\cal O}(\eps^0)$ is identified, which contributes at the level of squared amplitude. Further details are provided in App.~\ref{sec:details} and in Ref.~\cite{Cieri:2025xxx}, where we also analyze the double soft-gluon current at one loop in the SL region, with $p_1$ and $p_2$ in the physical initial state, and $p_1\parallel q_1$, $p_2\parallel q_2$, corresponding to Eq.~(\ref{eq:SLsoftcoll_fact_2sets}).
\section{Conclusion}
\label{sec:conclusion}
The soft and collinear factorization properties of hard-scattering amplitudes are crucial for understanding the quantum imprints of scattering processes and for developing the most sophisticated tools to achieve accurate theoretical predictions for relevant measurable quantities at high-energy colliders. In the limiting case with multiple soft partons, the scattering amplitude factorizes into a soft current, which controls the dominant singular behavior, and the reduced amplitude, which is derived from the original amplitude by simply removing the soft partons. As expected, due to the long-range interactions of the soft particles, the soft current depends on the momentum and color of all partons involved in the scattering process.
In the collinear limit, factorization is generally understood to be universal in the sense that the most singular factors, the splitting amplitudes, depend only on the momenta and degrees of freedom of the particles involved in the specific collinear splitting process. This property is called strict collinear factorization. The SCF is valid in the TL collinear region to all perturbative orders. However, as shown in the seminal paper~\cite{Catani:2011st}, the SCF breaks down for SL collinear configurations, for which a generalized factorization has been derived.
In this Letter, we have extended the collinear and simultaneous soft-collinear factorization to the most general kinematics with multiple collinear directions. For these configurations, which occur at high perturbative orders, naive multiplicative factorization does not hold in a general kinematics and the factorization needs to be further generalized, including the process dependence on the noncollinear partons. In the simultaneous soft-collinear limit, the scattering amplitudes do not factorize into a soft-collinear current and collinear splitting amplitudes. Instead, the factorization contains a generalized splitting amplitude that includes the effect of the dominant terms in the soft limit.
The SCF breaking at one loop for two collinear partons is anti-Hermitian~\cite{Catani:2011st} [see Eq.~(\ref{eq:j11gFB}) for a similar effect from the double-parton collinear limit of one loop single soft-gluon current] and therefore has no effect on the squared scattering amplitude. At higher multiplicities and higher orders, the SCF breaking effects are partially canceled also for squared amplitudes~\cite{Catani:2011st,Forshaw:2012bi,Henn:2024qjq,Becher:2024kmk,Becher:2025igg}. To the best of our knowledge, we have identified for the first time a factorization breaking term $\propto \pi^2$ at ${\cal O}(\eps^0)$ in the SL triple-collinear limit of one-loop double soft-gluon current, which contributes at the level of squared amplitude. Further details will be provided in Ref.~\cite{Cieri:2025xxx}.

\begin{acknowledgments}
We are grateful to Stefano Catani for many insightful discussions on the breaking of strict collinear factorization. We thank Roberto Bonciani and German Sborlini for useful discussions. This work is supported by the Spanish Government and ERDF/EU - Agencia Estatal de Investigaci\'on (MCIU/AEI/10.13039/501100011033), Grants No.PID2023-146220NB-I00, No. PID2020-114473GB-I00, and No. CEX2023-001292-S; and Generalitat Valenciana Grant No. PROMETEO/2021/071. L. C. is supported by Generalitat Valenciana GenT Excellence Programme (CIDEGENT/2020/011) and CSIC (ILINK22045). The work of P. K. D. is supported by European Commission MSCA Action COLLINEAR-FRACTURE, Grant Agreement No. 101108573.
\end{acknowledgments}
\appendix
\section{Additional details}
\label{sec:details}
For the convenience of readers, we provide additional details around the results presented in the core text of this letter, particularly in Eqs.~(\ref{eq:j12gFBR}) and (\ref{eq:Falp1alp2}). To this end, we first consider a single soft-gluon current at one loop~\cite{Catani:2000pi}, which is explicitly given by
\begin{align}
    \label{eq:j11g}
    \bom{J}^{(1)a}(q;h) &= -\gst^3 \frac{\cgt}{\eps^2} 
    \imath f^{a bc}\!\! \sum_{\substack{i,j\in h\\i\neq j}}\!\bom{T}_i^b \bom{T}_j^c \l[\fr{p_i\cdot \veps(q)}{p_i\cdot q} - \fr{p_j\cdot \veps(q)}{p_j\cdot q} \r]
    \nn\\
    &\times\frac{(-2p_i\cdot q - \imath 0)^{-\eps}(-2p_j\cdot q - \imath 0)^{-\eps}}{(-2p_i\cdot p_j - \imath 0)^{-\eps}},
\end{align}
where the soft gluon has momentum $q$, color index $a$, and polarization vector $\veps^{\mu}(q)$. The $f^{abc}$ is the SU($N$) structure constant, $\bom{T}_i$ is the SU($N$) color operator for parton~$i$. The $+\imath 0$ denotes the usual Feynman complex prescription, $\gst = g_{\rm S}\mu^{\eps}$, where $\eps$ is the dimensional regularization parameter in $d=4-2\eps$ spacetime dimensions, and
\begin{align}
    \label{eq:cgamma}
    \cgt = \cg \Gamma(1-\eps) \Gamma(1+\eps), \quad \cg = \fr{\Gamma(1+\eps)\Gamma^2(1-\eps)}{(4\pi)^{2-\eps}\Gamma(1-2\eps)}.
\end{align}

We now consider the double-parton collinear limit (say, $p_1\parallel q$) in both the TL ($p_1$ in the physical final state) and SL ($p_1$ and an additional hard parton $p_2$ both in the physical initial state) regions. We find the factorisation breaking (FB) contribution (see the core text for the notational details)
\begin{align}
    \label{eq:j11gFB}
    \sp^{(1)a}_{\rm FB}(q\subset c;\wp;h)\!&\eqv\!\sp^{(1)a}(q\subset c;\wp;h)\! - \!\sp^{(1)a}(q\subset c;\wp)
    \nn\\
    &\eqv \bom{J}^{(1)a}_{\rm SL}(q\subset c;\wp;h)-\bom{J}^{(1)a}_{\rm TL}(q\subset c;\wp)
    \nn\\
    &=-4\gst^3 f^{a bc}\bom{T}_1^b\bom{T}_2^c \frac{\cgt \sin(\pi\eps)}{\eps^2}
    \nn\\
    &\times \fr{p_1\cdot \veps(q)}{p_1\cdot q}(2p_1\cdot q)^{-\eps}\l(\fr{n\cdot q}{n\cdot p_1}\r)^{-\eps}.
\end{align}
In fact, the contribution in Eq.~(\ref{eq:j11gFB}) is anti-Hermitian, which means it cancels out in the squared amplitude.

Now, we proceed to examine the one-loop double soft-gluon current in the same spirit. To this end, we first separate its non-trivial part, $\Delta\boldsymbol{J}^{(1)a_1 a_2}_{\alpha_1\alpha_2}$\cite{Czakon:2022dwk,Cieri:2025xxx}, as follows:
\begin{align}
    \label{eq:j12gsep}
    \boldsymbol{J}^{(1)a_1 a_2}(q_1,q_2;h) &=  \Big[ \boldsymbol{J}^{(1)a_1 }_{\alpha_1}(q_1;h)\boldsymbol{J}^{(0)a_2}_{\alpha_2}(q_2;h)
    \nn\\
    &+ \boldsymbol{J}^{(1)a_2}_{\alpha_2}(q_2;h)\boldsymbol{J}^{(0)a_1}_{\alpha_1}(q_1;h) 
    \nn\\
    &+ \Delta\boldsymbol{J}^{(1)a_1 a_2}_{\alpha_1\alpha_2}(q_1,q_2;h)\Big]\veps_1^{\alpha_1}(q_1)\veps_2^{\alpha_2}(q_2).
\end{align}
Analogously, we write the corresponding tree-level double soft-gluon current~\cite{Czakon:2022dwk}.
\begin{align}
    \label{eq:j02gsep}
    \bom{J}^{(0)\,a_1 a_2}(q_1,q_2;h)
    &= \bigg[\fr{1}{2}\Big(\bom{{\hat J}}^{(0)a_1}_{\alpha_1}(q_1;h)\bom{J}^{(0)a_2}_{\alpha_2}(q_2;h)
    \nn\\
    &+\bom{{\hat J}}^{(0)a_2}_{\alpha_2}(q_2;h)\bom{J}^{(0)a_1}_{\alpha_1}(q_1);h\Big) 
    \nn\\
    &+ \Delta\bom{J}^{(0)a_1 a_2}_{\alpha_1\alpha_2}(q_1,q_2;h)\bigg]\veps_1^{\alpha_1}(q_1)\veps_2^{\alpha_2}(q_2),
\end{align}
where the tree-level single soft-gluon current is given by
\begin{align}
    \label{eq:j01g}
    \bom{J}^{(0)a}(q;h) &= \gst 
    \sum_{i\in h}\bom{T}_i^{a}\fr{p_i\cdot \veps(q)}{p_i\cdot q}.
\end{align}
To make the Ward identity explicit, it is re-expressed as
\begin{align}
    \label{eq:j01gAlt}
    \bom{{\hat J}}^{(0)a}(q;h) &= -\fr{\gst}{C_A}\imath f^{abc}\!\!\sum_{\substack{i,j\in h\\i\neq j}}\!\bom{T}_i^b \bom{T}_j^c \l( \fr{p_i\cdot \veps(q)}{p_i\cdot q}-\fr{p_j\cdot \veps(q)}{p_j\cdot q} \r)
    \nn\\
    &=\bom{J}^{(0)a}(q;h)\!-\!\fr{2\gst}{C_A}\imath f^{a bc}\!\sum_{\substack{i\in h}}\!\bom{T}_i^b \fr{p_i\cdot \veps(q)}{p_i\cdot q}\sum_{\substack{j\in h}}\!\bom{T}_j^c.
\end{align}

Similar to the single soft-gluon current at one-loop, we consider the triple-parton collinear limit $(p_1\parallel q_1\parallel q_2)$ in both the TL and SL regions. In this limit, we define
\begin{align}
    \label{eq:j12gFB}
    \Delta\sp^{(1)a_1 a_2}_{\rm FB}(q_1,q_2\subset c;\wp;h)&\eqv \Delta\bom{J}^{(1)a_1 a_2}_{\rm SL}(q_1,q_2\subset c;\wp;h)
    \nn\\
    &-\Delta\bom{J}^{(1)a_1 a_2}_{\rm TL}(q_1,q_2\subset c;\wp),
\end{align}
and the non-trivial part of the tree-level double soft-gluon current becomes
\begin{align}
    \label{eq:j02gNoFB}
    &\Delta\bom{J}^{(0)a_1 a_2}_{{\rm SL}}(q_1,q_2\subset c;\wp)=\Delta\bom{J}^{(0)a_1 a_2}_{{\rm TL}}(q_1,q_2\subset c;\wp)
    \nn\\
    &\simeq -\gst^2\fr{f^{a_1 bd}f^{a_2 cd}}{C_A}\bigg\{\bom{T}_1^b\bom{T}_1^c J^{(0)}_{\alpha_1\alpha_2}(q_1,q_2)
    \nn\\
    &+\bom{T}_1^c\bom{T}_1^b J^{(0)}_{\alpha_2\alpha_1}(q_2,q_1)\bigg\}\veps_1^{\alpha_1}(q_1)\veps_2^{\alpha_2}(q_2),
\end{align}
which defines the function $J^{(0)}_{\alpha_1\alpha_2}(q_1,q_2)$, and it is explicitly used to present the results in Eq.~(\ref{eq:j12gFBR}) in the core text of this letter.
\bibliographystyle{JHEP}
\bibliography{main}

\providecommand{\href}[2]{#2}\begingroup\raggedright\begin{thebibliography}{10}

\bibitem{Collins:1989gx}
J.~C. Collins, D.~E. Soper, and G.~F. Sterman, {\it {Factorization of Hard
  Processes in QCD}},  {\em Adv. Ser. Direct. High Energy Phys.} {\bf 5} (1989)
  1--91, [\href{http://arxiv.org/abs/hep-ph/0409313}{{\tt hep-ph/0409313}}].

\bibitem{Ellis:1996mzs}
R.~K. Ellis, W.~J. Stirling, and B.~R. Webber, {\em {QCD and collider
  physics}}, vol.~8.
\newblock Cambridge University Press, 2, 2011.

\bibitem{Altarelli:1977zs}
G.~Altarelli and G.~Parisi, {\it {Asymptotic Freedom in Parton Language}},
  {\em Nucl. Phys. B} {\bf 126} (1977) 298--318.

\bibitem{Bassetto:1983mvz}
A.~Bassetto, M.~Ciafaloni, and G.~Marchesini, {\it {Jet Structure and Infrared
  Sensitive Quantities in Perturbative QCD}},  {\em Phys. Rept.} {\bf 100}
  (1983) 201--272.

\bibitem{Frixione:1995ms}
S.~Frixione, Z.~Kunszt, and A.~Signer, {\it {Three jet cross-sections to
  next-to-leading order}},  {\em Nucl. Phys. B} {\bf 467} (1996) 399--442,
  [\href{http://arxiv.org/abs/hep-ph/9512328}{{\tt hep-ph/9512328}}].

\bibitem{Catani:1996vz}
S.~Catani and M.~H. Seymour, {\it {A General algorithm for calculating jet
  cross-sections in NLO QCD}},  {\em Nucl. Phys. B} {\bf 485} (1997) 291--419,
  [\href{http://arxiv.org/abs/hep-ph/9605323}{{\tt hep-ph/9605323}}]. [Erratum:
  Nucl.Phys.B 510, 503--504 (1998)].

\bibitem{Frixione:1997np}
S.~Frixione, {\it {A General approach to jet cross-sections in QCD}},  {\em
  Nucl. Phys. B} {\bf 507} (1997) 295--314,
  [\href{http://arxiv.org/abs/hep-ph/9706545}{{\tt hep-ph/9706545}}].

\bibitem{Catani:2002hc}
S.~Catani, S.~Dittmaier, M.~H. Seymour, and Z.~Trocsanyi, {\it {The Dipole
  formalism for next-to-leading order QCD calculations with massive partons}},
  {\em Nucl. Phys. B} {\bf 627} (2002) 189--265,
  [\href{http://arxiv.org/abs/hep-ph/0201036}{{\tt hep-ph/0201036}}].

\bibitem{Heinrich:2020ybq}
G.~Heinrich, {\it {Collider Physics at the Precision Frontier}},  {\em Phys.
  Rept.} {\bf 922} (2021) 1--69, [\href{http://arxiv.org/abs/2009.00516}{{\tt
  arXiv:2009.00516}}].

\bibitem{TorresBobadilla:2020ekr}
W.~J. Torres~Bobadilla et~al., {\it {May the four be with you: Novel
  IR-subtraction methods to tackle NNLO calculations}},  {\em Eur. Phys. J. C}
  {\bf 81} (2021), no.~3 250, [\href{http://arxiv.org/abs/2012.02567}{{\tt
  arXiv:2012.02567}}].

\bibitem{Agarwal:2021ais}
N.~Agarwal, L.~Magnea, C.~Signorile-Signorile, and A.~Tripathi, {\it {The
  infrared structure of perturbative gauge theories}},  {\em Phys. Rept.} {\bf
  994} (2023) 1--120, [\href{http://arxiv.org/abs/2112.07099}{{\tt
  arXiv:2112.07099}}].

\bibitem{Camarda:2021ict}
S.~Camarda, L.~Cieri, and G.~Ferrera, {\it {Drell\textendash{}Yan lepton-pair
  production: qT resummation at N3LL accuracy and fiducial cross sections at
  N3LO}},  {\em Phys. Rev. D} {\bf 104} (2021), no.~11 L111503,
  [\href{http://arxiv.org/abs/2103.04974}{{\tt arXiv:2103.04974}}].

\bibitem{Billis:2021ecs}
G.~Billis, B.~Dehnadi, M.~A. Ebert, J.~K.~L. Michel, and F.~J. Tackmann, {\it
  {Higgs pT Spectrum and Total Cross Section with Fiducial Cuts at Third
  Resummed and Fixed Order in QCD}},  {\em Phys. Rev. Lett.} {\bf 127} (2021),
  no.~7 072001, [\href{http://arxiv.org/abs/2102.08039}{{\tt
  arXiv:2102.08039}}].

\bibitem{Neumann:2022lft}
T.~Neumann and J.~Campbell, {\it {Fiducial Drell-Yan production at the LHC
  improved by transverse-momentum resummation at N4LLp+N3LO}},  {\em Phys. Rev.
  D} {\bf 107} (2023), no.~1 L011506,
  [\href{http://arxiv.org/abs/2207.07056}{{\tt arXiv:2207.07056}}].

\bibitem{Berends:1988zn}
F.~A. Berends and W.~T. Giele, {\it {Multiple Soft Gluon Radiation in Parton
  Processes}},  {\em Nucl. Phys. B} {\bf 313} (1989) 595--633.

\bibitem{Campbell:1997hg}
J.~M. Campbell and E.~W.~N. Glover, {\it {Double unresolved approximations to
  multiparton scattering amplitudes}},  {\em Nucl. Phys. B} {\bf 527} (1998)
  264--288, [\href{http://arxiv.org/abs/hep-ph/9710255}{{\tt hep-ph/9710255}}].

\bibitem{Catani:1999ss}
S.~Catani and M.~Grazzini, {\it {Infrared factorization of tree level QCD
  amplitudes at the next-to-next-to-leading order and beyond}},  {\em Nucl.
  Phys. B} {\bf 570} (2000) 287--325,
  [\href{http://arxiv.org/abs/hep-ph/9908523}{{\tt hep-ph/9908523}}].

\bibitem{Bern:1995ix}
Z.~Bern and G.~Chalmers, {\it {Factorization in one loop gauge theory}},  {\em
  Nucl. Phys. B} {\bf 447} (1995) 465--518,
  [\href{http://arxiv.org/abs/hep-ph/9503236}{{\tt hep-ph/9503236}}].

\bibitem{Bern:1998sc}
Z.~Bern, V.~Del~Duca, and C.~R. Schmidt, {\it {The Infrared behavior of one
  loop gluon amplitudes at next-to-next-to-leading order}},  {\em Phys. Lett.
  B} {\bf 445} (1998) 168--177,
  [\href{http://arxiv.org/abs/hep-ph/9810409}{{\tt hep-ph/9810409}}].

\bibitem{Bern:1999ry}
Z.~Bern, V.~Del~Duca, W.~B. Kilgore, and C.~R. Schmidt, {\it {The infrared
  behavior of one loop QCD amplitudes at next-to-next-to leading order}},  {\em
  Phys. Rev. D} {\bf 60} (1999) 116001,
  [\href{http://arxiv.org/abs/hep-ph/9903516}{{\tt hep-ph/9903516}}].

\bibitem{Catani:2000pi}
S.~Catani and M.~Grazzini, {\it {The soft gluon current at one loop order}},
  {\em Nucl. Phys. B} {\bf 591} (2000) 435--454,
  [\href{http://arxiv.org/abs/hep-ph/0007142}{{\tt hep-ph/0007142}}].

\bibitem{Bierenbaum:2011gg}
I.~Bierenbaum, M.~Czakon, and A.~Mitov, {\it {The singular behavior of one-loop
  massive QCD amplitudes with one external soft gluon}},  {\em Nucl. Phys. B}
  {\bf 856} (2012) 228--246, [\href{http://arxiv.org/abs/1107.4384}{{\tt
  arXiv:1107.4384}}].

\bibitem{Catani:2019nqv}
S.~Catani, D.~Colferai, and A.~Torrini, {\it {Triple (and quadruple) soft-gluon
  radiation in QCD hard scattering}},  {\em JHEP} {\bf 01} (2020) 118,
  [\href{http://arxiv.org/abs/1908.01616}{{\tt arXiv:1908.01616}}].

\bibitem{DelDuca:2022noh}
V.~Del~Duca, C.~Duhr, R.~Haindl, and Z.~Liu, {\it {Tree-level soft emission of
  a quark pair in association with a gluon}},  {\em JHEP} {\bf 01} (2023) 040,
  [\href{http://arxiv.org/abs/2206.01584}{{\tt arXiv:2206.01584}}].

\bibitem{Catani:2022hkb}
S.~Catani, L.~Cieri, D.~Colferai, and F.~Coradeschi, {\it {Soft
  gluon\textendash{}quark\textendash{}antiquark emission in QCD hard
  scattering}},  {\em Eur. Phys. J. C} {\bf 83} (2023), no.~1 38,
  [\href{http://arxiv.org/abs/2210.09397}{{\tt arXiv:2210.09397}}].

\bibitem{Zhu:2020ftr}
Y.~J. Zhu, {\it {Double soft current at one-loop in QCD}},  {\em JHEP} {\bf 02}
  (2026) 018, [\href{http://arxiv.org/abs/2009.08919}{{\tt arXiv:2009.08919}}].

\bibitem{Catani:2021kcy}
S.~Catani and L.~Cieri, {\it {Multiple soft radiation at one-loop order and the
  emission of a soft quark\textendash{}antiquark pair}},  {\em Eur. Phys. J. C}
  {\bf 82} (2022), no.~2 97, [\href{http://arxiv.org/abs/2108.13309}{{\tt
  arXiv:2108.13309}}].

\bibitem{Czakon:2022dwk}
M.~Czakon, F.~Eschment, and T.~Schellenberger, {\it {Revisiting the double-soft
  asymptotics of one-loop amplitudes in massless QCD}},  {\em JHEP} {\bf 04}
  (2023) 065, [\href{http://arxiv.org/abs/2211.06465}{{\tt arXiv:2211.06465}}].

\bibitem{Li:2013lsa}
Y.~Li and H.~X. Zhu, {\it {Single soft gluon emission at two loops}},  {\em
  JHEP} {\bf 11} (2013) 080, [\href{http://arxiv.org/abs/1309.4391}{{\tt
  arXiv:1309.4391}}].

\bibitem{Duhr:2013msa}
C.~Duhr and T.~Gehrmann, {\it {The two-loop soft current in dimensional
  regularization}},  {\em Phys. Lett. B} {\bf 727} (2013) 452--455,
  [\href{http://arxiv.org/abs/1309.4393}{{\tt arXiv:1309.4393}}].

\bibitem{Dixon:2019lnw}
L.~J. Dixon, E.~Herrmann, K.~Yan, and H.~X. Zhu, {\it {Soft gluon emission at
  two loops in full color}},  {\em JHEP} {\bf 05} (2020) 135,
  [\href{http://arxiv.org/abs/1912.09370}{{\tt arXiv:1912.09370}}].

\bibitem{Chen:2023hmk}
W.~Chen, M.-x. Luo, T.-Z. Yang, and H.~X. Zhu, {\it {Soft theorem to three
  loops in QCD and $ \mathcal{N} $ = 4 super Yang-Mills theory}},  {\em JHEP}
  {\bf 01} (2024) 131, [\href{http://arxiv.org/abs/2309.03832}{{\tt
  arXiv:2309.03832}}].

\bibitem{Herzog:2023sgb}
F.~Herzog, Y.~Ma, B.~Mistlberger, and A.~Suresh, {\it {Single-soft emissions
  for amplitudes with two colored particles at three loops}},  {\em JHEP} {\bf
  12} (2023) 023, [\href{http://arxiv.org/abs/2309.07884}{{\tt
  arXiv:2309.07884}}].

\bibitem{Chen:2024hvp}
X.~Chen and Z.~Liu, {\it {Tree-level soft emission for two pairs of quarks}},
  {\em JHEP} {\bf 02} (2025) 166, [\href{http://arxiv.org/abs/2411.08795}{{\tt
  arXiv:2411.08795}}].

\bibitem{Catani:1998nv}
S.~Catani and M.~Grazzini, {\it {Collinear factorization and splitting
  functions for next-to-next-to-leading order QCD calculations}},  {\em Phys.
  Lett. B} {\bf 446} (1999) 143--152,
  [\href{http://arxiv.org/abs/hep-ph/9810389}{{\tt hep-ph/9810389}}].

\bibitem{Dhani:2023uxu}
P.~K. Dhani, G.~Rodrigo, and G.~F.~R. Sborlini, {\it {Triple-collinear
  splittings with massive particles}},  {\em JHEP} {\bf 12} (2023) 188,
  [\href{http://arxiv.org/abs/2310.05803}{{\tt arXiv:2310.05803}}].

\bibitem{Craft:2023aew}
E.~Craft, M.~Gonzalez, K.~Lee, B.~Mecaj, and I.~Moult, {\it {The 1
  {\textrightarrow} 3 massive splitting functions from QCD factorization and
  SCET}},  {\em JHEP} {\bf 07} (2024) 080,
  [\href{http://arxiv.org/abs/2310.06736}{{\tt arXiv:2310.06736}}].

\bibitem{Kosower:1999rx}
D.~A. Kosower and P.~Uwer, {\it {One loop splitting amplitudes in gauge
  theory}},  {\em Nucl. Phys. B} {\bf 563} (1999) 477--505,
  [\href{http://arxiv.org/abs/hep-ph/9903515}{{\tt hep-ph/9903515}}].

\bibitem{Catani:2011st}
S.~Catani, D.~de~Florian, and G.~Rodrigo, {\it {Space-like (versus time-like)
  collinear limits in QCD: Is factorization violated?}},  {\em JHEP} {\bf 07}
  (2012) 026, [\href{http://arxiv.org/abs/1112.4405}{{\tt arXiv:1112.4405}}].

\bibitem{Sborlini:2013jba}
G.~F.~R. Sborlini, D.~de~Florian, and G.~Rodrigo, {\it {Double collinear
  splitting amplitudes at next-to-leading order}},  {\em JHEP} {\bf 01} (2014)
  018, [\href{http://arxiv.org/abs/1310.6841}{{\tt arXiv:1310.6841}}].

\bibitem{DelDuca:1999iql}
V.~Del~Duca, A.~Frizzo, and F.~Maltoni, {\it {Factorization of tree QCD
  amplitudes in the high-energy limit and in the collinear limit}},  {\em Nucl.
  Phys. B} {\bf 568} (2000) 211--262,
  [\href{http://arxiv.org/abs/hep-ph/9909464}{{\tt hep-ph/9909464}}].

\bibitem{Birthwright:2005ak}
T.~G. Birthwright, E.~W.~N. Glover, V.~V. Khoze, and P.~Marquard, {\it
  {Multi-gluon collinear limits from MHV diagrams}},  {\em JHEP} {\bf 05}
  (2005) 013, [\href{http://arxiv.org/abs/hep-ph/0503063}{{\tt
  hep-ph/0503063}}].

\bibitem{Birthwright:2005vi}
T.~G. Birthwright, E.~W.~N. Glover, V.~V. Khoze, and P.~Marquard, {\it
  {Collinear limits in QCD from MHV rules}},  {\em JHEP} {\bf 07} (2005) 068,
  [\href{http://arxiv.org/abs/hep-ph/0505219}{{\tt hep-ph/0505219}}].

\bibitem{DelDuca:2019ggv}
V.~Del~Duca, C.~Duhr, R.~Haindl, A.~Lazopoulos, and M.~Michel, {\it {Tree-level
  splitting amplitudes for a quark into four collinear partons}},  {\em JHEP}
  {\bf 02} (2020) 189, [\href{http://arxiv.org/abs/1912.06425}{{\tt
  arXiv:1912.06425}}].

\bibitem{DelDuca:2020vst}
V.~Del~Duca, C.~Duhr, R.~Haindl, A.~Lazopoulos, and M.~Michel, {\it {Tree-level
  splitting amplitudes for a gluon into four collinear partons}},  {\em JHEP}
  {\bf 10} (2020) 093, [\href{http://arxiv.org/abs/2007.05345}{{\tt
  arXiv:2007.05345}}].

\bibitem{Catani:2003vu}
S.~Catani, D.~de~Florian, and G.~Rodrigo, {\it {The Triple collinear limit of
  one loop QCD amplitudes}},  {\em Phys. Lett. B} {\bf 586} (2004) 323--331,
  [\href{http://arxiv.org/abs/hep-ph/0312067}{{\tt hep-ph/0312067}}].

\bibitem{Sborlini:2014mpa}
G.~F.~R. Sborlini, D.~de~Florian, and G.~Rodrigo, {\it {Triple collinear
  splitting functions at NLO for scattering processes with photons}},  {\em
  JHEP} {\bf 10} (2014) 161, [\href{http://arxiv.org/abs/1408.4821}{{\tt
  arXiv:1408.4821}}].

\bibitem{Sborlini:2014kla}
G.~F.~R. Sborlini, D.~de~Florian, and G.~Rodrigo, {\it {Polarized
  triple-collinear splitting functions at NLO for processes with photons}},
  {\em JHEP} {\bf 03} (2015) 021, [\href{http://arxiv.org/abs/1409.6137}{{\tt
  arXiv:1409.6137}}].

\bibitem{Badger:2015cxa}
S.~Badger, F.~Buciuni, and T.~Peraro, {\it {One-loop triple collinear splitting
  amplitudes in QCD}},  {\em JHEP} {\bf 09} (2015) 188,
  [\href{http://arxiv.org/abs/1507.05070}{{\tt arXiv:1507.05070}}].

\bibitem{Czakon:2022fqi}
M.~Czakon and S.~Sapeta, {\it {Complete collection of one-loop triple-collinear
  splitting operators for dimensionally-regulated QCD}},  {\em JHEP} {\bf 07}
  (2022) 052, [\href{http://arxiv.org/abs/2204.11801}{{\tt arXiv:2204.11801}}].

\bibitem{Bern:2004cz}
Z.~Bern, L.~J. Dixon, and D.~A. Kosower, {\it {Two-loop g ---\ensuremath{>} gg
  splitting amplitudes in QCD}},  {\em JHEP} {\bf 08} (2004) 012,
  [\href{http://arxiv.org/abs/hep-ph/0404293}{{\tt hep-ph/0404293}}].

\bibitem{Badger:2004uk}
S.~D. Badger and E.~W.~N. Glover, {\it {Two loop splitting functions in QCD}},
  {\em JHEP} {\bf 07} (2004) 040,
  [\href{http://arxiv.org/abs/hep-ph/0405236}{{\tt hep-ph/0405236}}].

\bibitem{Duhr:2014nda}
C.~Duhr, T.~Gehrmann, and M.~Jaquier, {\it {Two-loop splitting amplitudes and
  the single-real contribution to inclusive Higgs production at N$^3$LO}},
  {\em JHEP} {\bf 02} (2015) 077, [\href{http://arxiv.org/abs/1411.3587}{{\tt
  arXiv:1411.3587}}].

\bibitem{Guan:2024hlf}
X.~Guan, F.~Herzog, Y.~Ma, B.~Mistlberger, and A.~Suresh, {\it {Splitting
  amplitudes at N$^{3}$LO in QCD}},  {\em JHEP} {\bf 01} (2025) 090,
  [\href{http://arxiv.org/abs/2408.03019}{{\tt arXiv:2408.03019}}].

\bibitem{Rogers:2010dm}
T.~C. Rogers and P.~J. Mulders, {\it {No Generalized TMD-Factorization in
  Hadro-Production of High Transverse Momentum Hadrons}},  {\em Phys. Rev. D}
  {\bf 81} (2010) 094006, [\href{http://arxiv.org/abs/1001.2977}{{\tt
  arXiv:1001.2977}}].

\bibitem{Echevarria:2011epo}
M.~G. Echevarria, A.~Idilbi, and I.~Scimemi, {\it {Factorization Theorem For
  Drell-Yan At Low $q_T$ And Transverse Momentum Distributions
  On-The-Light-Cone}},  {\em JHEP} {\bf 07} (2012) 002,
  [\href{http://arxiv.org/abs/1111.4996}{{\tt arXiv:1111.4996}}].

\bibitem{Echevarria:2012js}
M.~G. Echevarr\'\i{}a, A.~Idilbi, and I.~Scimemi, {\it {Soft and Collinear
  Factorization and Transverse Momentum Dependent Parton Distribution
  Functions}},  {\em Phys. Lett. B} {\bf 726} (2013) 795--801,
  [\href{http://arxiv.org/abs/1211.1947}{{\tt arXiv:1211.1947}}].

\bibitem{Rogers:2013zha}
T.~C. Rogers, {\it {Extra spin asymmetries from the breakdown of
  transverse-momentum-dependent factorization in hadron-hadron collisions}},
  {\em Phys. Rev. D} {\bf 88} (2013), no.~1 014002,
  [\href{http://arxiv.org/abs/1304.4251}{{\tt arXiv:1304.4251}}].

\bibitem{Forshaw:2006fk}
J.~R. Forshaw, A.~Kyrieleis, and M.~H. Seymour, {\it {Super-leading logarithms
  in non-global observables in QCD}},  {\em JHEP} {\bf 08} (2006) 059,
  [\href{http://arxiv.org/abs/hep-ph/0604094}{{\tt hep-ph/0604094}}].

\bibitem{Forshaw:2008cq}
J.~R. Forshaw, A.~Kyrieleis, and M.~H. Seymour, {\it {Super-leading logarithms
  in non-global observables in QCD: Colour basis independent calculation}},
  {\em JHEP} {\bf 09} (2008) 128, [\href{http://arxiv.org/abs/0808.1269}{{\tt
  arXiv:0808.1269}}].

\bibitem{Keates:2009dn}
J.~Keates and M.~H. Seymour, {\it {Super-leading logarithms in non-global
  observables in QCD: Fixed order calculation}},  {\em JHEP} {\bf 04} (2009)
  040, [\href{http://arxiv.org/abs/0902.0477}{{\tt arXiv:0902.0477}}].

\bibitem{Becher:2021zkk}
T.~Becher, M.~Neubert, and D.~Y. Shao, {\it {Resummation of Super-Leading
  Logarithms}},  {\em Phys. Rev. Lett.} {\bf 127} (2021), no.~21 212002,
  [\href{http://arxiv.org/abs/2107.01212}{{\tt arXiv:2107.01212}}].

\bibitem{Becher:2024nqc}
T.~Becher, P.~Hager, G.~Martinelli, M.~Neubert, D.~Schwienbacher, and
  M.~Stillger, {\it {Super-leading logarithms in pp {\textrightarrow} 2 jets}},
   {\em JHEP} {\bf 01} (2025) 171, [\href{http://arxiv.org/abs/2411.12742}{{\tt
  arXiv:2411.12742}}].

\bibitem{Ma:2023gir}
Y.~Ma, G.~Sterman, and A.~Venkata, {\it {Soft Photon Theorem in QCD with
  Massless Quarks}},  {\em Phys. Rev. Lett.} {\bf 132} (2024), no.~9 091902,
  [\href{http://arxiv.org/abs/2311.06912}{{\tt arXiv:2311.06912}}].

\bibitem{Feige:2014wja}
I.~Feige and M.~D. Schwartz, {\it {Hard-Soft-Collinear Factorization to All
  Orders}},  {\em Phys. Rev. D} {\bf 90} (2014), no.~10 105020,
  [\href{http://arxiv.org/abs/1403.6472}{{\tt arXiv:1403.6472}}].

\bibitem{Forshaw:2012bi}
J.~R. Forshaw, M.~H. Seymour, and A.~Siodmok, {\it {On the Breaking of
  Collinear Factorization in QCD}},  {\em JHEP} {\bf 11} (2012) 066,
  [\href{http://arxiv.org/abs/1206.6363}{{\tt arXiv:1206.6363}}].

\bibitem{Sterman:2022gyf}
G.~Sterman, {\it {Comments on collinear factorization}},  in {\em {Snowmass
  2021}}, 7, 2022.
\newblock \href{http://arxiv.org/abs/2207.06507}{{\tt arXiv:2207.06507}}.

\bibitem{Duhr:2025lyg}
C.~Duhr, A.~Venkata, and C.~Zhang, {\it {Double Spacelike Collinear Limits from
  Multi-Regge Kinematics}},  {\em Phys. Rev. Lett.} {\bf 135} (2025), no.~24
  241601, [\href{http://arxiv.org/abs/2507.05355}{{\tt arXiv:2507.05355}}].

\bibitem{Cieri:2025xxx}
L.~Cieri, P.~K. Dhani, and G.~Rodrigo (to be published).

\bibitem{Henn:2024qjq}
J.~Henn, R.~Ma, Y.~Xu, K.~Yan, Y.~Zhang, and H.~X. Zhu, {\it {Two-loop
  spacelike splitting amplitude for N=4 Super-Yang-Mills theory}},  {\em Phys.
  Rev. D} {\bf 112} (2025), no.~7 076003,
  [\href{http://arxiv.org/abs/2406.14604}{{\tt arXiv:2406.14604}}].

\bibitem{Becher:2024kmk}
T.~Becher, P.~Hager, S.~Jaskiewicz, M.~Neubert, and D.~Schwienbacher, {\it
  {Factorization Restoration through Glauber Gluons}},  {\em Phys. Rev. Lett.}
  {\bf 134} (2025), no.~6 061901, [\href{http://arxiv.org/abs/2408.10308}{{\tt
  arXiv:2408.10308}}].

\bibitem{Becher:2025igg}
T.~Becher, P.~Hager, S.~Jaskiewicz, M.~Neubert, and D.~Schwienbacher, {\it
  {Low-energy theory of jet processes and PDF factorization}},  {\em JHEP} {\bf
  01} (2026) 024, [\href{http://arxiv.org/abs/2509.07082}{{\tt
  arXiv:2509.07082}}].

\end{thebibliography}\endgroup
\end{document}